\documentclass[twocolumn,showpacs,preprintnumbers,aps]{revtex4-1}

\usepackage{graphicx,color}

\usepackage{dcolumn}

\usepackage{bm}

\begin{document}

\preprint{}

\title{Frustrated magnetic interactions in an $S$=3/2 bilayer honeycomb lattice compound Bi$_3$Mn$_4$O$_{12}$(NO$_3$)}

\author{M. Matsuda}

\affiliation{Neutron Scattering Division, Oak Ridge National Laboratory, Oak Ridge, Tennessee 37831, USA}

\author{S. E. Dissanayake}

\altaffiliation[Present address: ]{Department of Physics, Duke University, Durham, North Carolina, 27708, USA}
\affiliation{Neutron Scattering Division, Oak Ridge National Laboratory, Oak Ridge, Tennessee 37831, USA}

\author{D. L. Abernathy}

\affiliation{Neutron Scattering Division, Oak Ridge National Laboratory, Oak Ridge, Tennessee 37831, USA}

\author{Y. Qiu}

\affiliation{NIST Center for Neutron Research, National Institute of Standards and Technology, Gaithersburg, Maryland 20899, USA}

\author{J. R. D. Copley}

\affiliation{NIST Center for Neutron Research, National Institute of Standards and Technology, Gaithersburg, Maryland 20899, USA}

\author{N. Kumada}

\affiliation{Center for Crystal Science and Technology, University of Yamanashi, 7-32 Miyamae, Kofu 400-8511, Japan}

\author{M. Azuma}

\affiliation{Laboratory for Materials and Structures, Tokyo Institute of Technology, 4259 Nagatsuta, Midori, Yokohama, 226-8503, Japan}

\date{\today}

\begin{abstract}
Inelastic neutron scattering study has been performed in an $S$=3/2 bilayer honeycomb lattice compound Bi$_3$Mn$_4$O$_{12}$(NO$_3$) at ambient and high magnetic fields. Relatively broad and monotonically dispersive magnetic excitations were observed at ambient field, where no long range magnetic order exists.
In the magnetic field-induced long-range ordered state at 10 T, the magnetic dispersions become slightly more intense, albeit still broad as in the disordered state, and two excitation gaps, probably originating from an easy-plane magnetic anisotropy and intrabilayer interactions, 
develop.
Analyzing the magnetic dispersions using the linear spin-wave theory, we estimated the intraplane and intrabilayer magnetic interactions, which are almost consistent with those determined by $ab$ $initio$ density functional theory calculations [M. Alaei $et$ $al.$, Phys. Rev. B {\bf 96}, 140404(R) (2017)], except the 3rd and 4th neighbor intrabilayer interactions. Most importantly, as predicted by the theory, there is no significant frustration in the honeycomb plane but frustrating intrabilayer interactions probably give rise to the disordered ground state.
\end{abstract}


\maketitle

\section{Introduction}
Intensive studies have been performed both experimentally and theoretically in the Ising spin system on the two-dimensional (2D) honeycomb lattice with Kitaev model \cite{Kitaev}, in which three different inequivalent nearest-neighbor interactions exist, since this system exhibits an exotic spin liquid ground state characterized by a topological order. On the other hand, conventional 2D honeycomb lattice magnets, in which the nearest-neighbor interaction is dominant, are unfrustrated. However, frustration can be induced in the presence of second-neighbor interactions, which gives rise to interesting new phases \cite{takano,katsura,Mulder,Okumura,Wang,Mosadeq,Clark,Lu,Ganesh1,Zhang,Rosales,Ganesh,Zhu1,Zhu2}, such as spin-liquid states.

\begin{figure}
\includegraphics[width=8.5cm]{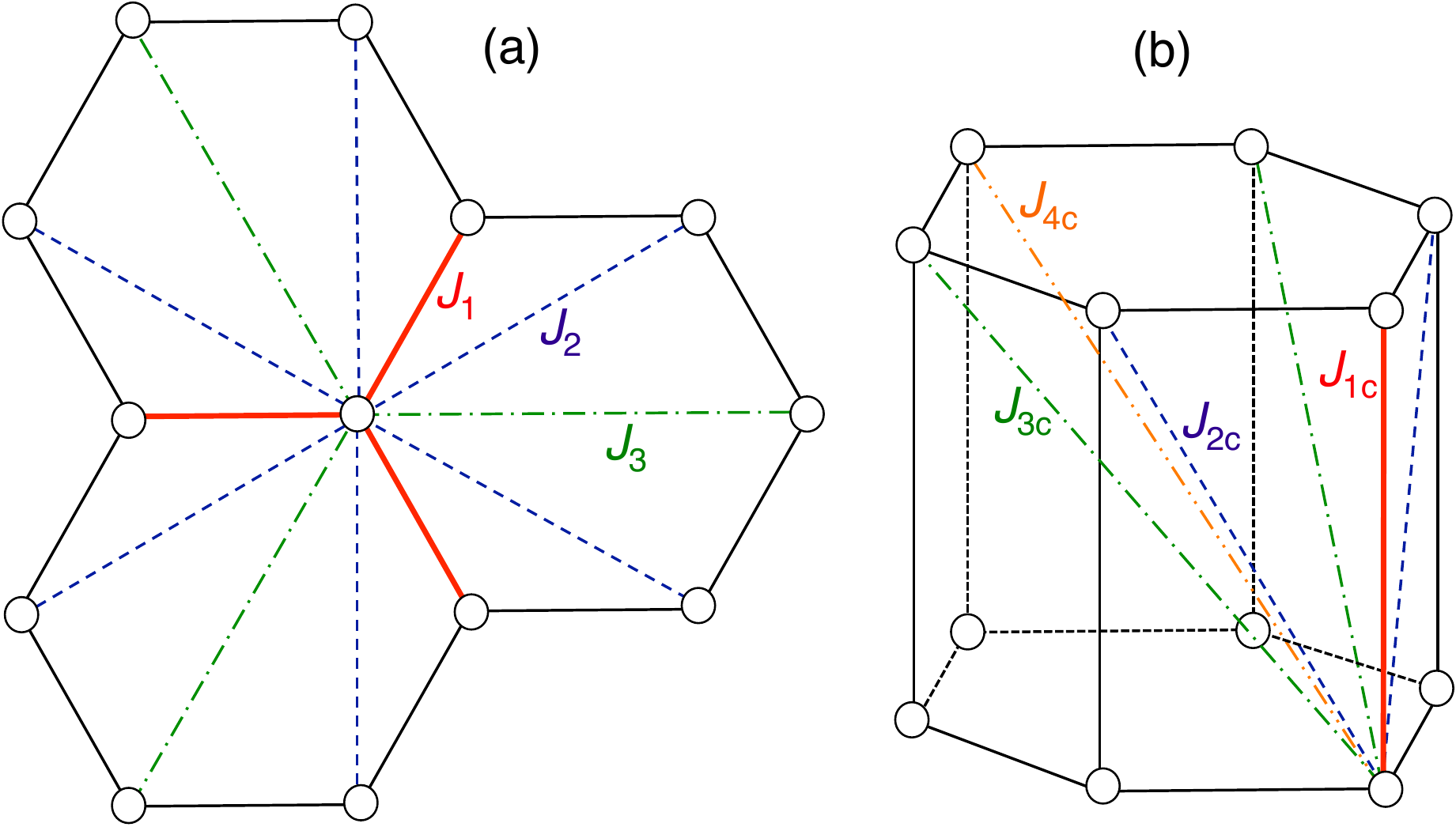}
\caption{(Color online) Schematic structure of the honeycomb bilayer in Bi$_3$Mn$_4$O$_{12}$(NO$_3$). The open circles represent the Mn$^{4+}$ moments. Magnetic interactions in a honeycomb plane $J_1$ (bold solid line), $J_2$ (dashed line), and $J_3$ (long dashed dotted line) (a) and intrabilayer interactions $J_{1c}$ (bold solid line), $J_{2c}$ (dashed line), $J_{3c}$ (long dashed dotted line), and $J_{4c}$ (long dashed double-dotted line) (b), which are used to calculate the spin-wave dispersions.}
\label{structure}
\end{figure}

Bi$_3$Mn$_4$O$_{12}$(NO$_3$) (BMNO), which has a trigonal structure ($P$3), consists of undistorted bilayer honeycomb lattices of the magnetic Mn$^{4+}$ ions (nominally $t_{2g}^3; S = 3/2)$ without orbital degree of freedom~\cite{azuma_JACS}, as shown in Fig. \ref{structure}. The magnetic susceptibility shows a broad maximum centered around 70 K, indicating a characteristic feature of two-dimensional antiferromagnets.
The Curie-Weiss temperature has a large value of $\Theta_{CW}$ = $-$257 K~\cite{azuma_JACS} and $-$222 K~\cite{azuma}, indicating overall strong antiferromagnetic interactions. The material does not show a long range magnetic order down to at least 0.4 K much lower than $|\Theta_{CW}|$, indicative of the presence of strong frustration~\cite{azuma_JACS}. Pure powder sample was found to exhibit a spin-glass behavior below 6 K \cite{azuma}.
Matsuda $et$ $al.$ performed neutron diffraction measurements in BMNO and found that it shows a short-range antiferromagnetic order at low temperatures \cite{Matsuda}.

With applying magnetic field, a long range antiferromagnetic order (LRAFO) phase appears around 6 T \cite{Matsuda}. The magnetic structure in this magnetic-field induced phase has an antiparallel arrangement between nearest-neighbor spins in the honeycomb plane with the spins pointing along a direction in the plane and an antiparallel arrangement between nearest-neighbor spins along the $c$ axis.
The low critical magnetic field suggests that the disordered ground state and the magnetic-field induced state are energetically close.
The intrabilayer couplings were suggested to be important to explain the disordered ground state \cite{Kandpal,wadati}. It was reported that an large intrabilayer antiferromagnetic nearest-neighbor coupling gives rise to dimerization \cite{Ganesh2,Oitmaa,Zhang1,Zhang2}. On the other hand, the frustrating further-neighbor interactions beyond those of second-neighbors both inplane and intrabilayer were suggested to cause the disordered state \cite{azuma,wadati,Albuquerque,Cabra,Bishop,Gomez}. Although these further-neighbor interactions should be important, the bilayer structure with many relevant interactions makes the detailed analysis of the inelastic neutron scattering results challenging.

\begin{table*}
\caption{Magnetic interactions in Bi$_3$Mn$_4$O$_{12}$(NO$_3$) determined by the linear spin-wave analysis on the inelastic neutron scattering results and by the DFT calculations \cite{Alaei}. The difference of the absolute values between the experimental and theoretical study originates from the difference in the definition of the spin magnitude. Our linear spin-wave calculations were performed for the spin Hamiltonian [Eq. (1)] with $S$=3/2, whereas Alaei $et$ $al.$ used classical unit vectors in their spin Hamiltonian. The interactions normalized to $J_1$ are also shown to compare the experimental and theoretical results.
$D$ represents the single-ion anisotropy.}
\begin{ruledtabular}
\begin{tabular}{cccccccccc}
Method & $J_1$ & $J_2$ & $J_3$ & $J_{1c}$ & $J_{2c}$ & $J_{3c}$ & $J_{4c}$ & $D$ & $\Theta_{CW}$\\
\hline
Neutron (this work) & 3.3 meV & 0.46 meV & 0.29 meV & 0.96 meV & 0.30 meV & $-$ & $-$ & 0.012 meV & $-$215.2 K\\
normalized & 1 & 0.14 & 0.088 & 0.29 & 0.09 & $-$ & $-$ & \\
DFT (Ref. \cite{Alaei})      & 10.7 meV & 0.9 meV & 1.2 meV & 3.0 meV & 1.1 meV & 0.5 meV & 0.9 meV & $-$ & $-$244 K\\
normalized  & 1 & 0.084 & 0.112 & 0.28 & 0.10 & 0.047 & 0.084 & $-$\\
\end{tabular}
\end{ruledtabular}
\end{table*}
Recently, Alaei $et$ $al.$ reported a mechanism of the frustrating interactions in BMNO obtained using the $ab$ $initio$ density functional theory (DFT) calculations \cite{Alaei}. They calculated three intraplane interactions, $J_1$, $J_2$, and $J_3$ [Fig. \ref{structure}(a)], which represent nearest, second-nearest, and third-nearest neighbor's interactions in the honeycomb plane, respectively and four intrabilayer interactions $J_{1c}$, $J_{2c}$, $J_{3c}$, and $J_{4c}$ [Fig. \ref{structure}(b)], which represent intrabilayer nearest, next-nearest, third-, and fourth-neighbor's interactions, respectively. The calculated interactions are shown in Table I. All the interactions are calculated to be antiferromagnetic with $J_1$ largest and $J_{1c}$ second largest. $J_2$ could compete with $J_1$ in the honeycomb plane. If only $J_1$ and $J_2$ are relevant, the magnetic ground state becomes disordered when $J_2/J_1 >$ 0.15 for $S$=3/2~\cite{takano}. However, since $J_2$ is much smaller than $J_1$, a collinear N\'{e}el order is preferred. On the other hand, intrabilayer couplings $J_{1c}$ and $J_{2c}$ were found to compete strongly. This frustration could make the effective intrabilayer interaction negligibly small, which makes the magnetic interactions purely two-dimensional, where no long-range magnetic order develops at finite temperatures. As a result, LRAFO is predicted to be suppressed in a narrow range of $J_{2c}$. With $J_{2c}$ smaller than a critical value, the intrabilayer spin arrangement between the nearest-neighbor spins becomes antiparallel due to $J_{1c}$ being dominant, which is actually realized in BMNO. On the other hand, with $J_{2c}$ larger than the critical value, the intrabilayer spin arrangement becomes parallel due to $J_{2c}$ being dominant. The calculated value of $J_{2c}$ for BMNO is very close to the critical value.

To elucidate the frustration mechanism in BMNO, we performed inelastic neutron scattering experiments using a powder sample of BMNO at ambient and high magnetic fields and analyzed the observed magnetic dispersions using the linear spin-wave theory. Since the magnetic excitations are dispersive monotonically and not so sharp even in the field-induced LRAFO state, it was challenging to evaluate the predicted seven magnetic interactions accurately. Fortunately, we found a few characteristics of the magnetic dispersions which can be used to estimate the magnetic interactions.
We confirmed that the magnetic interactions ($J_1$, $J_2$, $J_3$, $J_{1c}$, and $J_{2c}$) predicted theoretically almost reproduce the observed magnetic dispersions, indicating that the disordered ground state is driven by the frustrating $J_{1c}$ and $J_{2c}$.

\section{Experimental Details}
A powder sample of BMNO was prepared by hydrothermal synthesis \cite{azuma_JACS}. The powder sample that weighs $\sim$4 g was used. Although the sample contains 6.7 weight \% of MnO$_2$, as shown in Ref. \cite{azuma_JACS}, inelastic scattering from the impurity phase was negligibly small \cite{impurity}.
The inelastic neutron scattering experiments were carried out on a disk chopper spectrometer (DCS) \cite{DCS} installed at the NIST Center for Neutron Research (NCNR) and a chopper neutron spectrometer ARCS \cite{ARCS} installed at the Spallation Neutron Source (SNS) at Oak Ridge National Laboratory (ORNL). We utilized two incident energies of 6.3 and 25.3 meV on DCS and an incident energy of 20.0 meV on ARCS. Energy resolutions at the elastic position are $\sim$0.4 and $\sim$1.5 meV with $E\rm_i$ = 6.3 and 25.3 meV, respectively, on DCS and $\sim$0.7 meV with $E\rm_i$ = 20.0 meV on ARCS. The vertical magnetic field was applied up to 10 T using a superconducting magnet on DCS. The magnetic field dependence of the magnetic dispersions were measured at 2.1 K on DCS. Temperature dependence of the magnetic dispersions were measured at ambient field in the temperature range of 5 $\le T \le$ 250 K using a closed-cycle refrigerator on ARCS.
The visualization of the inelastic neutron scattering data were performed using the DAVE software \cite{DAVE}.

\section{Results and Discussion}
\subsection{Inelastic Neutron Scattering}
\subsubsection{Ambient Magnetic Field}
\begin{figure}
\includegraphics[width=8.2cm]{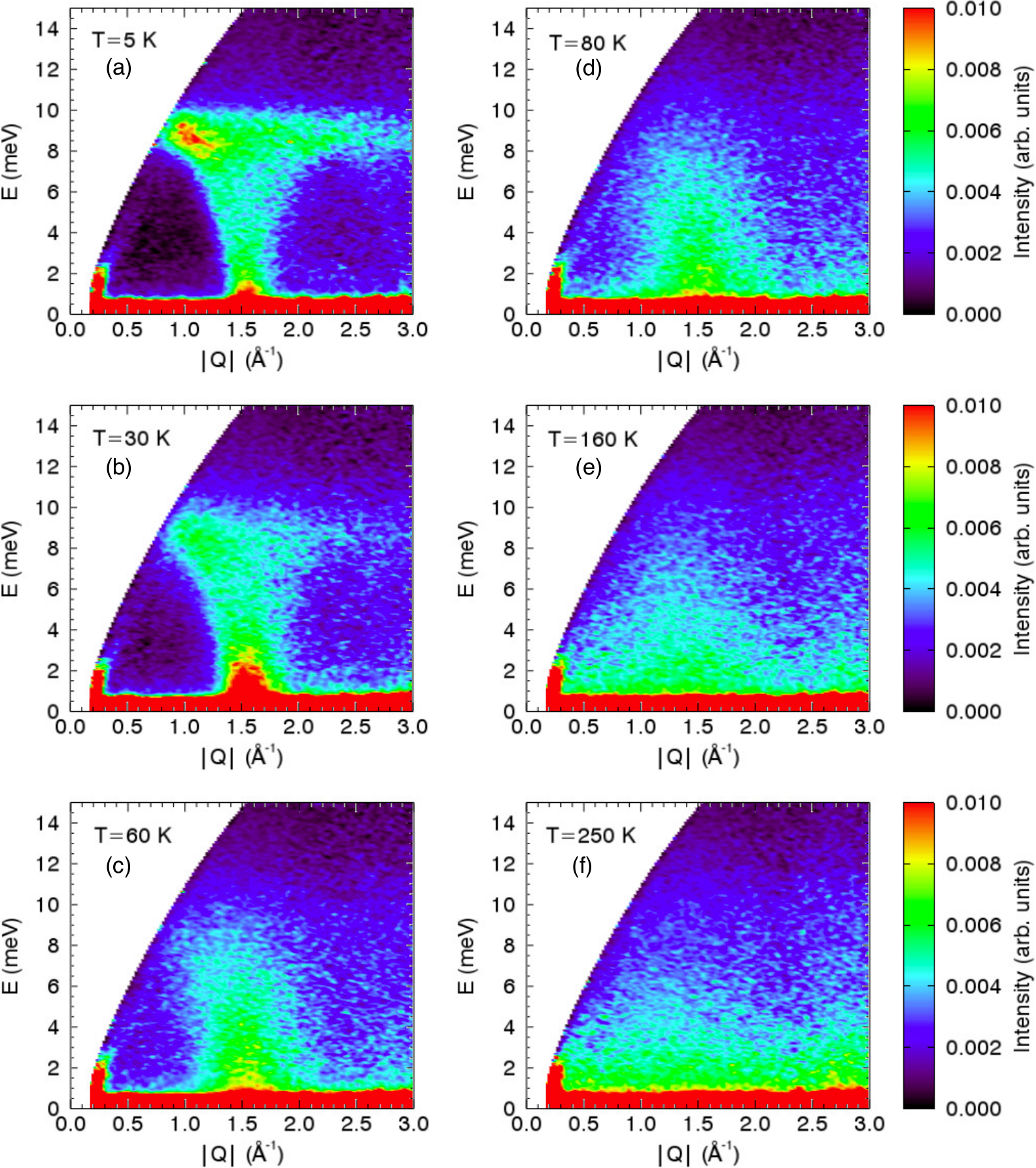}
\caption{(Color online) Color contour maps of the inelastic neutron scattering intensity $S(|Q|,E)$ for Bi$_3$Mn$_4$O$_{12}$(NO$_3$) powder measured with $E\rm_i$=20.0 meV on ARCS at ambient field and at $T$=5 (a), 30 (b), 60 (c), 80 (d), 160 (e), and 250 K (f).}
\label{ARCS}
\end{figure}
\begin{figure}
\includegraphics[width=8.5cm]{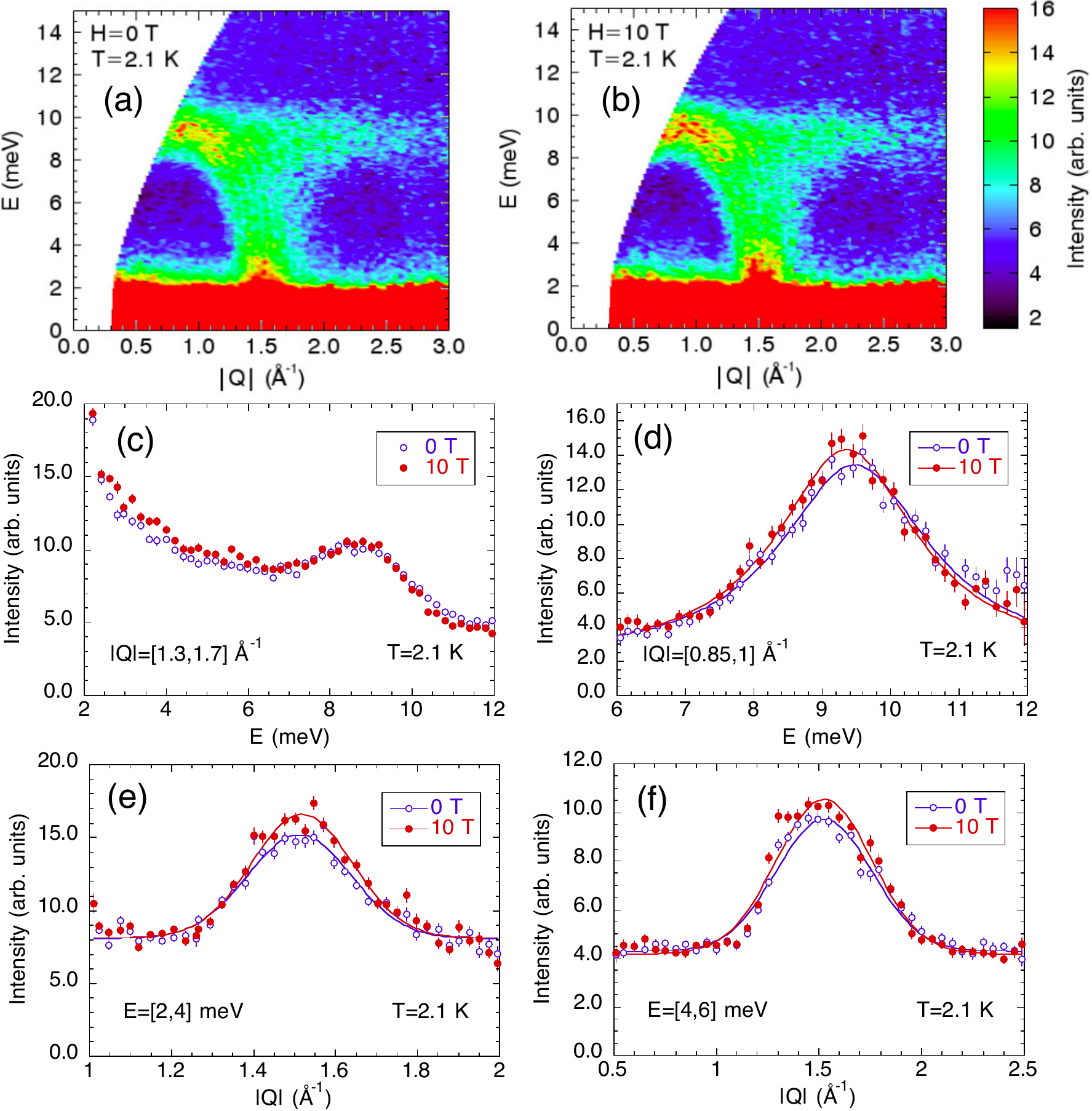}
\caption{(Color online) Color contour maps of $S(|Q|,E)$ for Bi$_3$Mn$_4$O$_{12}$(NO$_3$) powder measured with $E\rm_i$=25.3 meV on DCS at $T$=2.1 K and at $H$=0 (a) and 10 T (b). Energy-cuts of the excitations at $H$=0 and 10 T integrated over the range of 1.3$\le Q\le$1.7 \AA$^{-1}$ (c) and 0.85$\le Q\le$1 \AA$^{-1}$ (d). $Q$-cuts of the excitations at $H$=0 and 10 T integrated over the range of 2$\le E\le$4 meV (e) and 4$\le E\le$6 meV (f). Solid lines in (d) are fits with a Lorentzian function. Solid lines in (e) and (f) are fits with a Gaussian function.}
\label{DCS_HE}
\end{figure}
Figure \ref{ARCS}(a) shows an image plot of inelastic neutron spectrum measured on ARCS at $T$=5 K and $H$=0 T. Although there is no LRAFO, well dispersive magnetic excitations were observed. The spin-wave-like excitation rises from $Q\sim$1.5 \AA$^{-1}$, corresponding to the (101) Bragg peak position, and the band width of the excitation is $\sim$9.5 meV. The observed excitations are consistent with those reported in Ref. \cite{Matsuda}, which shows the results measured with coarser energy and $Q$ resolutions. An excitation gap originating from the magnetic anisotropy was not observed.
With increasing temperature, the magnetic excitations become broader, as shown in Fig. \ref{ARCS}(b)-(f). At 60 K, the dispersive mode almost disappears and a broad excitation centered around 1.5 \AA$^{-1}$ remains, which originates from short-range antiferromagnetic correlations. The excitation becomes broader in $Q$ with increasing temperature, since the correlation length becomes shorter at higher temperatures. The excitations become close to those from paramagnetic state at 250 K. The temperature dependence is consistent with that of the magnetic susceptibility which has a large $\Theta_{CW}$ of $-$257 or $-$222 K. As described in Sec. III-B. the magnetic excitations at low temperatures can be analyzed using the linear spin-wave theory and the estimated magnetic interactions reproduce $\Theta_{CW}$.

\subsubsection{Magnetic Field}
BMNO exhibits a LRAFO under magnetic field above 6 T at 1.5 K, as reported in Ref. \cite{Matsuda}.
Magnetic excitations measured in the field-induced phase with $E\rm_i$=25.3 meV using a superconducting magnet on DCS are shown in Fig. \ref{DCS_HE}, which shows the overall feature up to 15 meV, as in Fig. \ref{ARCS}. The ambient field data in Fig. \ref{DCS_HE}(a) is consistent with that in Fig. \ref{ARCS}(a). Figure \ref{DCS_HE}(b) shows the magnetic excitations at 10 T, where the LRAFO is observed. Figure \ref{DCS_HE}(c) shows an energy-cut of the magnetic excitations around 1.5 \AA$^{-1}$, where the magnetic zone center is located. The excitations are slightly more intense at 10 T than at 0 T below $\sim$6 meV. This is consistent with the $Q$-cut intensities shown in Figs. \ref{DCS_HE}(e) and (f).
On the other hand, the intensities at 0 and 10 T are similar in the range of 6$\le E\le$10 meV and the intensity at 10 T is slightly less intense than at 0 T around 11 meV.
The intensity in the range of 8$\le E\le$9.5 meV at low-$Q$ below 1 \AA$^{-1}$ is enhanced at 10 T, as shown in Fig. \ref{DCS_HE}(d).
Therefore, the overall magnetic excitation signal is enhanced at 10 T, although the enhancement is not homogeneous in the energy and momentum space. We can also conclude that there is almost no perceivable change in the energy scale and the sharpness of the dispersions.

\begin{figure}
\includegraphics[width=8.5cm]{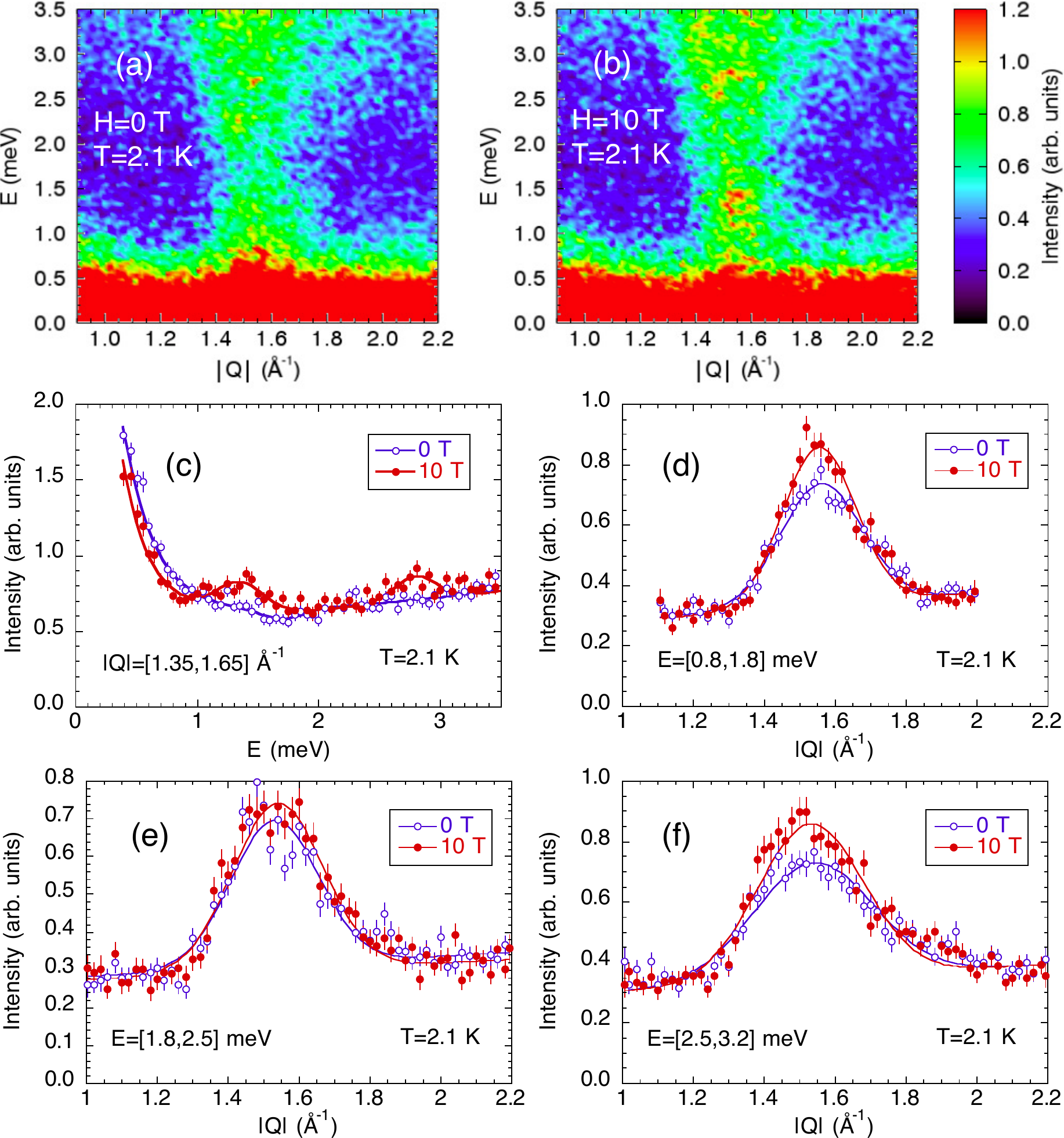}
\caption{(Color online) Low energy magnetic excitations in Bi$_3$Mn$_4$O$_{12}$(NO$_3$) powder measured at $T$=2.1 K with $E\rm_i$=6.3 meV on DCS. Color contour maps of $S(|Q|,E)$ at $H$=0 (a) and 10 T (b). (c) Energy-cut of the excitations integrated over the range of 1.4$\le Q\le$1.65 \AA$^{-1}$ at $H$=0 and 10 T. $Q$-cut of the excitations at $H$=0 and 10 T integrated over the range of 0.8$\le E\le$1.8 meV (d), 1.8$\le E\le$2.5 meV (e), and 2.5$\le E\le$3.2 meV (f). Solid lines in (c) are guides to the eye. Solid lines in (d), (e), and (f) are fits with a Gaussian function.}
\label{DCS_LE}
\end{figure}
In order to elucidate the excitation width and excitation gaps in detail,
we measured low-energy magnetic excitations with a high energy resolution ($E\rm_i$=6.3 meV). Figures \ref{DCS_LE}(a) and (b) show magnetic excitations below 3.5 meV at 0 and 10 T, respectively. It is visible that the intensities around 1.3 and 2.7 meV are enhanced at 10 T but those below 1 meV is reduced. This feature is also seen in the energy-cut plot in Fig. \ref{DCS_LE}(c). The $Q$-cut plots in Figs. \ref{DCS_LE}(d), (e), and (f) also confirm an enhancement of the intensity at the two energies and no enhancement between them at 10 T. Due to the enhanced signals two peaks appear at the two energies [Fig. \ref{DCS_LE}(c)], suggesting that these peaks are bottom of gapped excitations.
As described in Sec III-B, our analysis indicates that the lower and higher energy gaps originate from the easy-plane anisotropy in the honeycomb plane and the intrabilayer couplings, respectively.
As shown in Fig. \ref{DCS_LE}(e), the excitation width of $Q$ does not change at 10 T, which is consistent with the results measured with $E\rm_i$=25.3 meV.

A possible origin for the small difference between magnetic excitations at 0 and 10 T is that some finite fraction of the disordered phase still remains at 10 T~\cite{Matsuda}. It was interpreted that the field-induced transition depends on magnetic field direction and only the magnetic moments perpendicular to the magnetic field show the long-range order and the others remain disordered. Due to the mixture of the two phases, the magnetic excitations at 10 T can be still broadened.
Another possible origin is that the intraplane and intrabilayer magnetic correlations might be sufficient to generate reasonably well-defined magnetic excitations at zero magnetic field. The correlation lengths are estimated to be $\sim$8 \AA\ intraplane and $\sim$6 \AA\ intrabilayer \cite{Matsuda}, which contain several hexagons. The correlation lengths might be underestimated so that more hexagons could be correlated. This effect sharpens the magnetic excitations at 0 T, which can be similar to those at 10 T.

\subsection{Spin-Wave Calculations}
\begin{figure*}
\includegraphics[width=18cm]{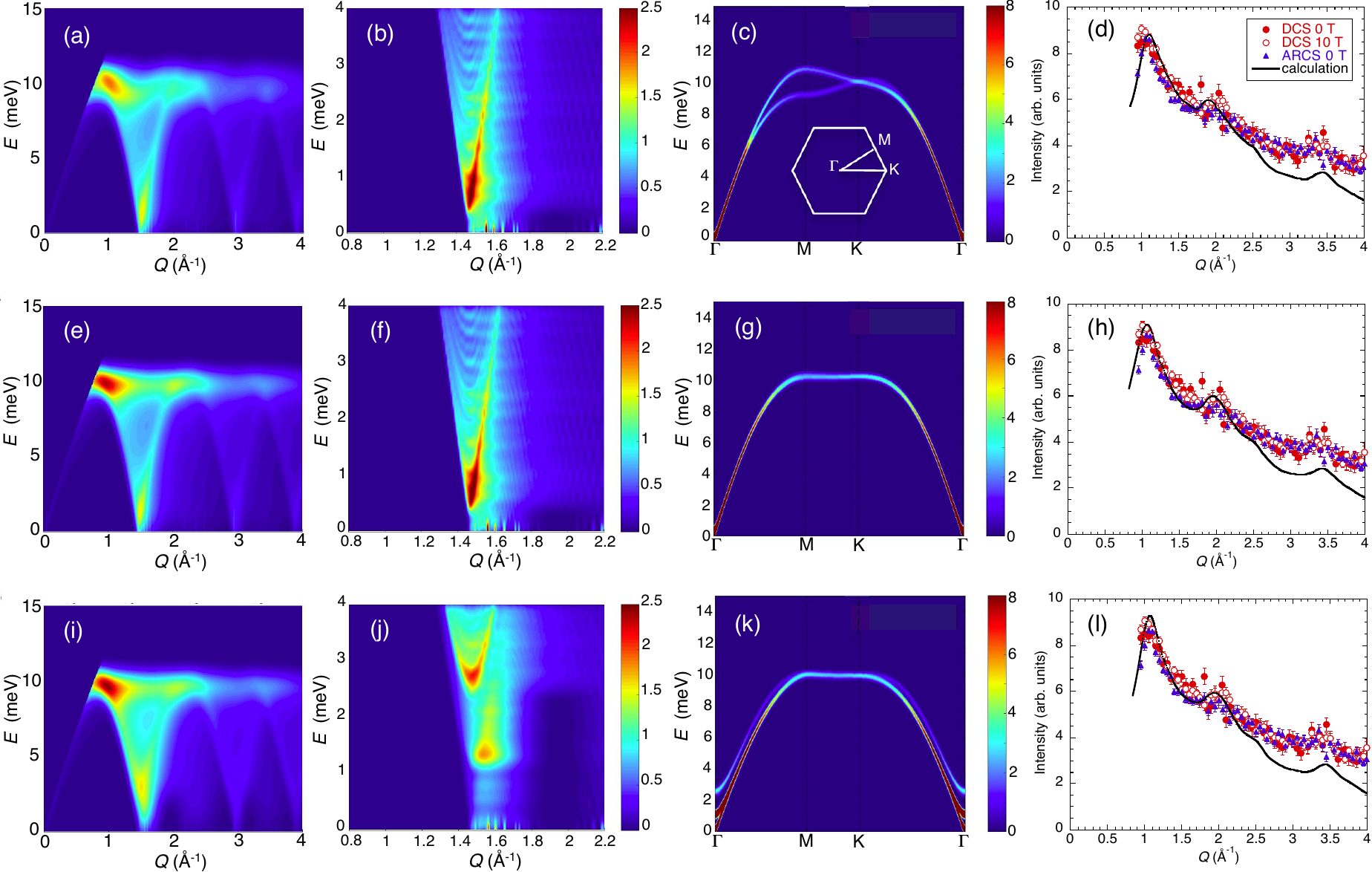}
\caption{(Color online) Summary of the calculated magnetic excitations using the linear spin-wave theory with magnetic field (10 T) perpendicular to the spin direction. (a), (e), and (i) Powder averaged inelastic neutron scattering intensity $S(|Q|,E)$. (b), (f), and (j)  Low-energy and low-$Q$ region of $S(|Q|,E)$. (c), (g), and (k) Magnetic dispersions along the high symmetric directions in the reciprocal space. (d), (h), and (l) $Q$ dependence of the scattering intensities $S(|Q|,E)$ integrated over the energy range of 8$\le E\le$11 meV. The experimental values measured on ARCS at 0 T and 5 K and on DCS at 0 and 10 T and 2.1 K are also shown. (a)-(d) are the results with theoretically predicted interactions $J_2$=0.084$J_1$, $J_3$=0.112$J_1$, $J_{1c}$=0.28$J_1$, $J_{2c}$=0.10$J_1$, $J_{3c}$=0.047$J_1$, and $J_{4c}$=0.084$J_1$. $J_1$=2.64 meV is used to adjust the energy scale of the experimental data.
(e)-(h) are the results with $J_1$=2.55 meV, $J_2$=0.084$J_1$, $J_3$=0.112$J_1$, $J_{1c}$=0.28$J_1$, and $J_{2c}$=0.094$J_1$. Note that $J_{2c}$ is fixed at a slightly smaller value to keep the ground state stable with the spin correlations same as those observed. (i)-(l) are the results with $J_1$=3.3 meV, $J_2$=0.14$J_1$, $J_3$=0.088$J_1$, $J_{1c}$=0.29$J_1$, $J_{2c}$=0.09$J_1$, and $D$=0.012 meV.}
\label{SpinW}
\end{figure*}
In order to determine the magnetic interactions from the observed magnetic dispersions, we performed linear spin-wave calculations for the spin Hamiltonian of the bilayer Heisenberg honeycomb lattice, which is described by 
\begin{equation}
H=\sum_{i>j}J_{ij}{\mathbf{S}}_{i}\cdot {\mathbf{S}}_{j} + \sum_{i} D S_{z,i}^2 - g\mu_B\sum_{i}{\mathbf{S}_{i}}\cdot {\mathbf{H}},
\label{Hamiltonian}
\end{equation}

where $J_{ij}$ represents the intralayer and intrabilayer magnetic interactions shown in Fig. \ref{structure}. The second term is the single-ion anisotropy term. The spins are reported to point along a direction in the honeycomb plane in the field-induced LRAFO phase \cite{Matsuda}. Therefore, $D$ is supposed to have a positive value. The third term is the Zeeman term, where $g$, $\mu_B$, and ${\mathbf{H}}$ represent $g$-factor, the Bohr magneton, and magnetic field, respectively.

SpinW package \cite{spinw} was used for the calculations. We calculated powder averaged magnetic dispersions using the linear spin-wave model. It is not straightforward to compare the experimental and calculated results because of the large number of the relevant interactions and the relatively broad and monotonic magnetic dispersions observed. After some trials, we found the following characteristics useful to analyze the magnetic dispersions. (1) The dispersion around the band maximum ($\sim$9.5 meV) is almost flat, indicating that the dispersion around the zone boundary should not have a strong structure. (2) $Q$ dependence of the intensities integrated over the energy range of 8$\le E\le$11 meV is sensitive to the interactions. (3) The excitation gaps at 1.3 and 2.7 meV should be reproduced. As shown in Sec. III A2, (3) is a clear difference between excitations at 0 and 10 T. This is probably because the gaps become broadened in the disordered phase. As described below, the gaps originate from the anisotropy and intrabilayer couplings. We tried to reproduce (1)-(3) with the model calculations. We confirmed that the interaction parameters used for the spin-wave calculations reproduce the observed antiferromagnetic structure in magnetic field.

We first make it clear how the magnetic field affects spin components parallel and perpendicular to the magnetic field. As in Eq. (1), the former fully affects the Hamiltonian and splits the spin-wave dispersions, whereas the latter has no effect and the dispersions are the same as in the zero magnetic field. The calculations with the magnetic field parallel to the spin direction are shown in Appendix. Since the measurements were performed with a powder sample, the magnetic field directions are random.
As mentioned in Sec. I, the disordered phase still remains at 10 T. Since about 1/3 of the broad signals remains, it was suggested in Ref. \cite{Matsuda} that the spin component with magnetic field effectively applied along the $c$ axis gives rise to the LRAFO and other spin component is disordered. This model is consistent with the magnetic field dependence of the magnetic excitations, which were actually observed, as follows. As shown in Sec. III A2, we did not observe any splits of the magnetic excitations at the top of the dispersions at 10 T. This indicates that the observed magnetic excitations at 10 T are not affected by the magnetic field. Therefore, the excitations probably originate from a mixture of the field-induced LRAFO, in which the spin components perpendicular to the magnetic field contribute, and disordered phases. The excitations from the disordered phase probably do not split in magnetic fields due to the short range correlations.

The magnetic dispersions were first calculated using a set of magnetic interactions predicted by Alaei $et$ $al.$ \cite{Alaei} (Table I). Note that we cannot directly compare the absolute values of the interactions obtained experimentally and theoretically, since the definition of the Hamiltonians are different. Therefore, $J_1$ was varied and the ratios for the other interactions were fixed in the calculation. The magnetic dispersions calculated in this way
are presented in Figs. \ref{SpinW}(a)-(d).
Although the $Q$ dependence of the energy-integrated intensity calculated agrees reasonably well with that observed experimentally \cite{note1} [Fig. \ref{SpinW}(d)], the calculation indicates an upturn of the dispersion at low-$Q$ [Fig. \ref{SpinW}(a)], which originates from the dispersive modes between $M$ and $K$ points [Fig. \ref{SpinW}(c)].
Although no anisotropy term is included, there exists a finite gap. This originates from the intrabilayer couplings, as reported in Ref. \cite{Zhang1}.
Figures \ref{SpinW}(e)-(h) show the results of calculations without $J_{3c}$ and $J_{4c}$. Now the upturn of the dispersion is absent, indicating that  $J_{3c}$ and $J_{4c}$ should be very small.
We evaluated $J_{1}$, $J_{2}$, $J_{3}$, $J_{1c}$, and $J_{2c}$ and found that all these are necessary to reproduce the dispersions.
In order to reproduce the observed two excitation gaps, the  anisotropy term should also be considered as well as the intrabilayer couplings.
We tried to calculate with various combinations of these interactions to reproduce the observed dispersions.
Our best estimated values are $J_{1}$=3.3 meV, $J_{2}$=0.46 meV(=0.14$J_{1}$), $J_{3}$=0.29 meV(=0.088$J_{1}$), $J_{1c}$=0.96 meV(=0.29$J_{1}$), $J_{2c}$=0.30 meV(=0.09$J_{1}$), and $D$=0.012 meV. The results of the calculation with these interactions are shown in Figs. \ref{SpinW}(i)-(l). All the features of the dispersions are reproduced reasonably well. $J_{2}$ and $J_{3}$ are slightly larger and smaller than those predicted theoretically, respectively. On the other hand, $J_{1c}$ and $J_{2c}$ are in excellent agreement with the predicted values. With these interactions, $\Theta_{CW}$ is calculated to be $-$215.2 K, which is close to the previously reported values ($-$257 and $-$222 K)~\cite{azuma_JACS,azuma}.

According to our calculations, the lower-energy gap originates from the easy-plane anisotropy. The higher-energy gap is driven by the intrabilayer interactions $J_{1c}$ and $J_{2c}$. $J_{1c}$ increases the gap energy, whereas $J_{2c}$ decreases the gap energy. $J_{1}$, $J_{2}$, and $J_{3}$ also affect the two gap energies. The experimentally estimated $J_{1c}$ and $J_{2c}$, which are similar to those theoretically predicted, give rise to reasonable gap energies.
A Dzyaloshinskii-Moriya (DM) interaction is suggested from the ESR results in Ref. \cite{Okubo}. However, our calculation including the DM interaction diverges and generates imaginary scattering intensity, indicating that the magnetic state becomes unstable. Therefore, the DM interaction is considered to be weak even if it exists.

\section{Summary}
Inelastic neutron scattering study has been performed in an $S$=3/2 bilayer honeycomb lattice compound BMNO to observe the magnetic excitations. The overall magnetic excitations do not change much between the ambient-field disordered and field-induced LRAFO phases. The magnetic dispersions measured in the LRAFO state are slightly more intense and two excitation gaps, probably originating from the single-ion anisotropy and the intrabilayer couplings, become clear.
The magnetic interactions estimated using the linear spin-wave theory are almost consistent with those determined by the DFT calculations, except $J_{3c}$ and $J_{4c}$. Our experimental result confirms that there is no significant frustration in the honeycomb plane. Instead, the frustrating intrabilayer interactions $J_{1c}$ and $J_{2c}$ probably give rise to the disordered ground state, as reported in Ref. \cite{Alaei}.

\begin{acknowledgments}
This research used resources at the Spallation Neutron Source, a DOE Office of Science User Facility operated by the Oak Ridge National Laboratory. Access to DCS was provided by the Center for High Resolution Neutron Scattering, a partnership between the National Institute of Standards and Technology and the National Science Foundation under Agreement No. DMR-1508249. This work was partially supported by the World Research Hub Initiative (WRHI) of Tokyo Institute of Technology.
\end{acknowledgments}

\appendix*
\newcommand{\hbAppendixPrefix}{A}
\renewcommand{\thefigure}{\hbAppendixPrefix\arabic{figure}}
\setcounter{figure}{0}
\renewcommand{\thetable}{\hbAppendixPrefix\arabic{table}} 
\setcounter{table}{0}
\renewcommand{\theequation}{\hbAppendixPrefix\arabic{equation}} 
\setcounter{equation}{0}
    \section{Spin-Wave Dispersions in Magnetic fields Parallel and Perpendicular to the Spin Direction}
\begin{figure*}
\includegraphics[width=18cm]{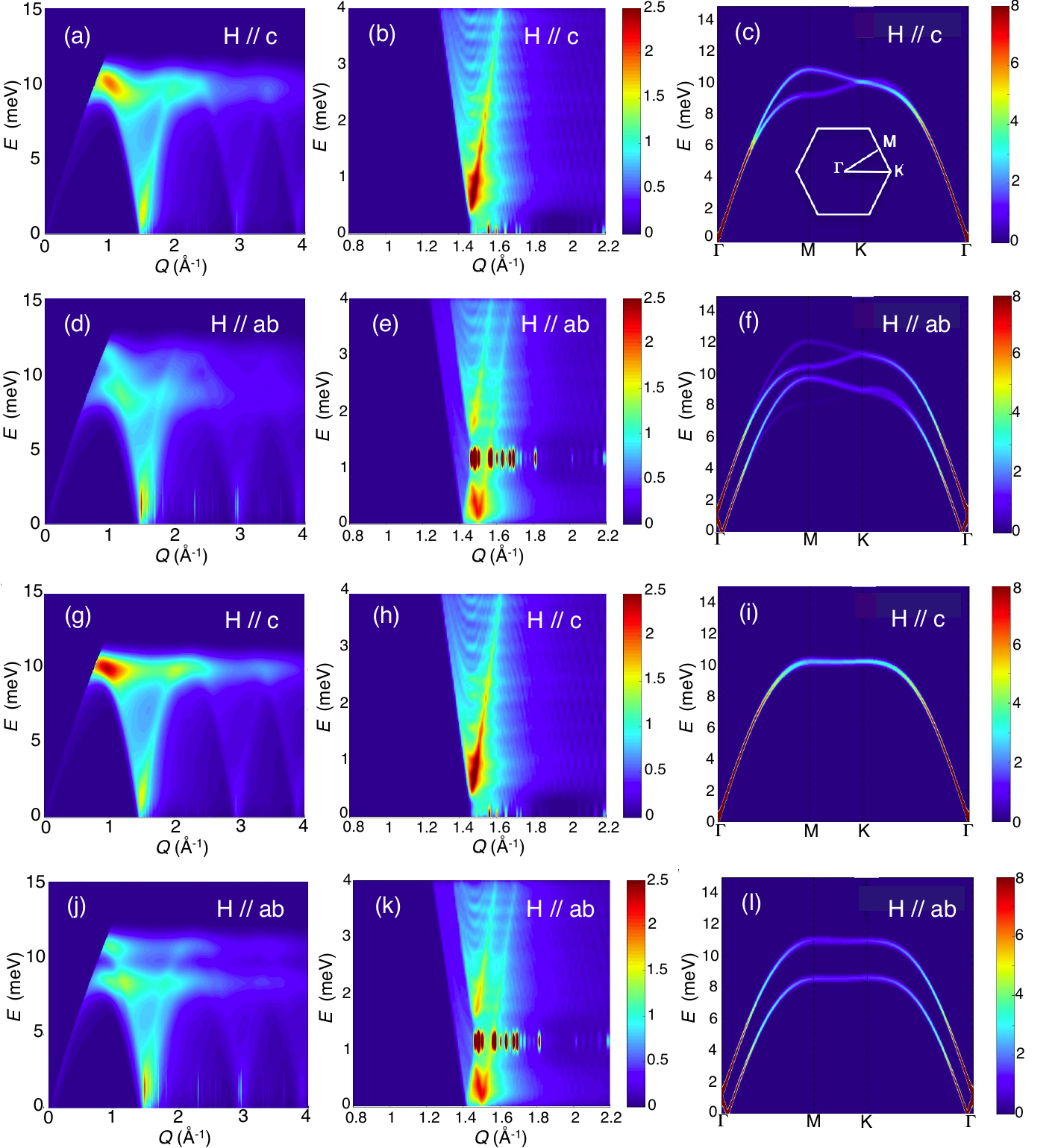}
\caption{(Color online) Summary of the calculated magnetic excitations in the magnetic field of 10 T using the linear spin-wave theory. (a), (d), (g), and (j) Powder averaged inelastic neutron scattering intensity $S(|Q|,E)$. (b), (e), (h), and (k)  Low-energy and low-$Q$ region of $S(|Q|,E)$. (c), (f), (i) and (l) Magnetic dispersions along the high symmetric directions in the reciprocal space. (a)-(f) are the results with $J_1$=2.64 meV, $J_2$=0.084$J_1$, $J_3$=0.112$J_1$, $J_{1c}$=0.28$J_1$, $J_{2c}$=0.10$J_1$, $J_{3c}$=0.047$J_1$, and $J_{4c}$=0.084$J_1$. The ratios of the interactions theoretically predicted \cite{Alaei} are used.
[(a)-(c) Magnetic field perpendicular to the spins. (d)-(f) Magnetic field parallel to the spins.] (g)-(l) are the results with $J_1$=2.55 meV, $J_2$=0.084$J_1$, $J_3$=0.112$J_1$, $J_{1c}$=0.28$J_1$, and $J_{2c}$=0.093$J_1$.
[(g)-(i) Magnetic field perpendicular to the spins. (j)-(l) Magnetic field parallel to the spins.]
The sharp and vertically elongated spots seen at $\sim$1.2 meV in (e) and (k) originate from the sharp edge of the dispersion at $\Gamma$ point, generated by the longitudinal magnetic field. Due to the almost flat dispersion along the $c$ axis, many spots are present at the same energy.}
\label{SpinW_field}
\end{figure*}

The magnetic excitations are calculated using the spin Hamiltonian [Eq. (1)]. For the Zeeman term, the magnetic field affects the spin-wave dispersions differently, depending on whether it is applied along or perpendicular to the spin direction. Fig. \ref{SpinW_field} displays the magnetic dispersions with magnetic field perpendicular and parallel to the spin direction. Although the dispersions with magnetic fields perpendicular to the spin direction are similar to those experimentally observed, those with field parallel to the spin direction are very different. In particular, the split of the dispersion modes at the top of the excitation band was not observed experimentally. These results indicate that the magnetic field effectively applied perpendicular to the $c$ axis gives rise to the LRAFO and the excitations we observed at 10 T are from a mixture of the field-induced LRAFO and disordered phases. The latter is not probably affected by the magnetic field of 10 T due to the short-range correlations. This picture of the magnetic phases in magnetic field is quite consistent with that suggested by the neutron diffraction study in magnetic field \cite{Matsuda}.

\begin{thebibliography}{}
\bibitem{Kitaev}A. Kitaev, Ann. Phys. (N. Y.), {\bf 321}, 2 (2006).
\bibitem{takano}K. Takano, Phys. Rev. B {\bf 74}, 140402(R) (2006).
\bibitem{katsura}S. Katsura, T. Ide, and T. Morita, J. Stat. Phys. {\bf 74}, 381 (1986).
\bibitem{Mulder}A. Mulder, R. Ganesh, L. Capriotti, and A. Paramekanti, Phys. Rev. B {\bf 81}, 214419 (2010).
\bibitem{Okumura}S. Okumura, H. Kawamura, T. Okubo, and Y. Motome, J. Phys. Soc. Jpn. {\bf 79}, 114705 (2010).
\bibitem{Wang}F. Wang, Phys. Rev. B {\bf 82}, 024419 (2010).
\bibitem{Mosadeq}H. Mosadeq, F. Shahbazi, and S. A. Jafari, J. Phys. Condens. Matter {\bf 23}, 226006 (2011).
\bibitem{Clark}B. K. Clark, D. A. Abanin, and S. L. Sondhi, Phys. Rev. Lett. {\bf 107}, 087204 (2011).
\bibitem{Lu}Y.-M. Lu and Y. Ran, Phys. Rev. B {\bf 84}, 024420 (2011).
\bibitem{Ganesh1}R. Ganesh, D. N. Sheng, Y.-J. Kim, and A. Paramekanti, Phys. Rev. B {\bf 83}, 144414 (2011).
\bibitem{Zhang}H. Zhang and C. A. Lamas, Phys. Rev. B {\bf 87}, 024415 (2013).
\bibitem{Rosales}H. D. Rosales, D. C. Cabra, C. A. Lamas, P. Pujol, and M. E. Zhitomirsky, Phys. Rev. B {\bf 87}, 104402 (2013).
\bibitem{Ganesh}R. Ganesh, J. van den Brink, and S. Nishimoto, Phys. Rev. Lett. {\bf 110}, 127203 (2013).
\bibitem{Zhu1}Z. Zhu, D. A. Huse, and S. R. White, Phys. Rev. Lett. {\bf 110}, 127205 (2013).
\bibitem{Zhu2}Z. Zhu, D. A. Huse, and S. R. White, Phys. Rev. Lett. {\bf 111}, 257201 (2013).
\bibitem{azuma_JACS}O. Smirnova, M. Azuma, N. Kumada, Y. Kusano, M. Matsuda, Y. Shimakawa, T. Takei, Y. Yonesaki, and N. Kinomura, J. Am. Chem. Soc. {\bf 131}, 8313 (2009).
\bibitem{impurity}The impurity phase MnO$_2$ shows an incommensurate antiferromagnetic order below $\sim$93 K with an ordered moment of $\sim$2.34$\mu\rm_B$ \cite{MnO2,MnO2_2}, similar to that in BMNO. There has been no report of the magnetic excitation measurement in this material. The amount of the impurity phase is tiny so that the inelastic intensity should be minor. If the impurity phase affect the excitations, there should be well-defined spin-wave excitations at 60 K, where the magnetic Bragg peak intensity reaches to $\sim$75\% of the saturated value \cite{MnO2}. However, as shown in Fig. 2(c), only broad excitations were observed, which indicates that the observed signal is mostly from BMNO.
\bibitem{azuma}N. Onishi, K. Oka, M. Azuma, Y. Shimakawa, Y. Motome, T. Taniguchi, M. Hiraishi, M. Miyazaki, T. Masuda, A. Koda, K. M. Kojima, and R. Kadono, Phys. Rev. B {\bf 85}, 184412 (2012).
\bibitem{Matsuda}M. Matsuda, M. Azuma, M. Tokunaga, Y. Shimakawa, and N. Kumada, Phys. Rev. Lett. {\bf 105}, 187201 (2010).
\bibitem{Kandpal}H. C. Kandpal and J. van den Brink, Phys. Rev. B {\bf 83}, 140412(R) (2011).
\bibitem{wadati}H. Wadati, K. Kato, Y. Wakisaka, T. Sudayama, D. G. Hawthorn, T. Z. Regier, N. Onishi, M. Azuma, Y. Shimakawa, T. Mizokawa, A. Tanaka, G. A. Sawatzky, Solid State. Commun. {\bf 162}, 18 (2013).
\bibitem{Oitmaa}J. Oitmaa and R. R. P. Singh, Phys. Rev. B {\bf 85}, 014428 (2012).
\bibitem{Zhang1}H. Zhang, M. Arlego, and C. A. Lamas, Phys. Rev. B {\bf 89}, 024403 (2014).
\bibitem{Zhang2}H. Zhang, C. A. Lamas, M. Arlego, and W. Brenig, Phys. Rev. B {\bf 93}, 235150 (2016).
\bibitem{Ganesh2}R. Ganesh, S. V. Isakov, and A. Paramekanti, Phys. Rev. B {\bf 84}, 214412 (2011).
\bibitem{Albuquerque}A. F. Albuquerque, D. Schwandt, B. Het\'{e}nyi, S. Capponi, M. Mambrini, and A. M. La\"{u}chli, Phys. Rev. B {\bf 84}, 024406 (2011).
\bibitem{Cabra}D. C. Cabra, C. A. Lamas, and H. D. Rosales, Phys. Rev. B {\bf 83}, 094506 (2011).
\bibitem{Bishop}R. F. Bishop and P. H. Y. Li, Phys. Rev. B {\bf 85}, 155135 (2012).
\bibitem{Gomez}F. A. G\'{o}mez Albarrac\'{i}n, and H. D. Rosales, Phys. Rev. B {\bf 93}, 144413 (2016).
\bibitem{Alaei}M. Alaei, H. Mosadeq, I. A. Sarsari, and F. Shahbazi, Phys. Rev. B {\bf 96}, 140404(R) (2017).
\bibitem{DCS}J. R. D. Copley and J. C. Cook, Chem. Phys. {\bf 292}, 477 (2003).
\bibitem{ARCS}D. L. Abernathy, M. B. Stone, M. J. Loguillo, M. S. Lucas, O. Delaire, X. Tang, J. Y. Y. Lin, and B. Fultz, Rev. Sci. Instrum. {\bf 83}, 015114 (2012).
\bibitem{DAVE}R. T. Azuah, L. R. Kneller, Y. Qiu, P. L. W. Tregenna-Piggott, C. M. Brown, J. R. D. Copley, and R. M. Dimeo, J. Res. Natl. Inst. Stan. Technol. {\bf 114}, 341 (2009).
\bibitem{spinw}S. Toth and B. Lake, J. Phys.: Condens. Matter {\bf 27}, 166002 (2015).
\bibitem{note1}The calculated intensities become smaller than those observed at the $Q$ region above $\sim$2.5 \AA$^{-1}$. This is probably because the magnetic form factor of the free Mn$^{4+}$ moment used for the calculation may deviate from the actual value in BMNO at high $Q$. For example, it was reported in YBa$_2$Cu$_3$O$_{6.15}$ that the observed magnetic intensities in the first Brillouin zone is larger than the calculated values with free Cu$^{2+}$ moment. This is ascribed to an anisotropic distribution of the moment, that is, the moments lie in the $d_{x^2-y^2}$ orbital \cite{shamoto}. The magnetic form factor in BMNO, where moments are present in $d_{xy}$, $d_{yz}$, and $d_{xz}$ orbitals, may also be affected.
\bibitem{Okubo}S. Okubo, T. Ueda, H. Ohta, W. Zhang, T. Sakurai, N. Onishi, M. Azuma, Y. Shimakawa, H. Nakano, and T. Sakai, Phys. Rev. B {\bf 86}, 140401 (2012).
\bibitem{MnO2}H. Sato, K. Wakiya, T. Enoki, T. Kiyama, Y. Wakabayashi, H. Nakao, and Y. Murakami, J. Phys. Soc. Jpn., {\bf 70}, 37 (2001).
\bibitem{MnO2_2}M. Regulski, R. Przenios\l{}o, I. Sosnowska, and J.-U. Hoffmann, J. Phys. Soc. Jpn., {\bf 73}, 3444 (2004).
\bibitem{shamoto}S. Shamoto, M. Sato, J. M. Tranquada, B. J. Sternlieb, and G. Shirane, Phys. Rev. B {\bf 48}, 13817 (1993).
\end{thebibliography}
\end{document}